\documentclass[reprint,amsmath,amssymb,pre,floatfix]{revtex4} 
\usepackage{graphicx}

\begin{document}

\title{Boundaries on Neutrino Mass from Supernovae Neutronization Burst by Liquid Argon Experiments}


\author{F.~Rossi-Torres}
\email{ftorres@ifi.unicamp.br}
\author{M.~M.~Guzzo}
\email{guzzo@ifi.unicamp.br}
\author{E.~Kemp}
\email{kemp@ifi.unicamp.br}
\affiliation{Instituto de F\'isica Gleb Wataghin, Universidade Estadual de Campinas - UNICAMP, Rua S\'ergio Buarque de Holanda, 777, 13083-859, Campinas-SP, Brazil}
\begin{abstract}
This work presents an upper bound on the neutrino mass using the 
emission of $\nu_e$ from the neutronization burst of a core 
collapsing supernova at 10~kpc of distance and a progenitor star of 
15~M$_\odot$. The calculations were done considering a 34 kton Liquid 
Argon Time Projection Chamber similar to the Far Detector proposal of 
the Long Baseline Neutrino Experiment (LBNE). We have performed a Monte 
Carlo simulation for the number of events integrated in 5~ms bins. Our 
results are $m_\nu<2.71$~eV and $0.18~\mbox{eV}<m_\nu<1.70$~eV, at 95\% 
C.L, assuming normal hierarchy and inverted hierarchy, respectively. We 
have analysed different configurations for the detector performance 
resulting in neutrino mass bound of $\mathcal{O}(1)$~eV.
\end{abstract}
\maketitle

\section{Introduction}
\label{intro}

From the current experimental evidence of neutrino 
oscillations~\cite{sno:osc_evidence,kamland:osc_evidence}, we know 
that neutrinos are massive particles, and each neutrino flavor 
eigenstate $|\nu_\beta\rangle \; (\beta=e,\mu,\tau)$ is a coherent superposition 
of mass eingenstates $|\nu_i\rangle \; (i=1,2,3)$, connected by the PMNS mixing matrix $U_{\beta i}$, through the relation  $|\nu_\beta\rangle =\sum_i U_{\beta i} |\nu_i\rangle$. However, oscillation 
experiments are sensitive only to the mass squared differences ($\Delta 
m^2_{ij}=m^2_i-m^2_j \; ; \; i,j=1,2,3$) and not on the absolute value 
of each mass eigenstate which is one of the most challenging measurement 
in particle physics.
Furthermore, one must be careful about the interpretation of experiments 
on neutrino masses since different techniques result in different 
observables. For instance, neutrino-less double beta decay experiments 
search for Majorana neutrinos with mass 
$m^2_{ee}=|\sum_{i}U^2_{ei}m_i|^2$ and single-$\beta$ decay experiments 
observe features driven by massive neutrinos with 
$m^2_e=\sum_{i}|U^2_{ei}|^2m^2_i$ not dependent on Majorana phases. 
Since $m_e > m_{ee}$ measurements which are not exclusively sensitive to 
Majorana neutrinos can be safely taken as conservative ones~\cite{AbsNuMassRev}. 


If we can measure the mass of any neutrino flavor $\nu_\beta$ the masses of 
$\nu_i$ can be obtained from the boundaries of the mixing matrix and 
other $\Delta m^2_{ij}$ obtained from different oscillation experiments.

The strongest experimental limit in the neutrino mass, as quoted in the 
Particle Data Group (PDG)~\cite{pdg}, was obtained by the 
Troitsk experiment on tritium single-$\beta$ decay: 
$m_{e}<$~2.05~eV at 95\% confidence level (C.L)~\cite{troitsk}. 
This kind of experiment measures the modification of the $\beta$ 
spectrum near its endpoint caused by the neutrino mass. The KATRIN 
experiment~\cite{katrin} will improve the technique and plan to set a 
stronger limit by enhancing the sensitivity at 
least in one order of magnitude. 




Neutrinos emitted from stellar collapse, for example type II supernovae, can give a valuable contribution to the determination of the mass of neutrinos, as idealized by Zatsepin~\cite{zatsepin}. Using neutrino data~\cite{kii,imb,baksan} from the SN1987A, which exploded in the Large Magellanic cloud ($\sim$50~kpc), and the likelihood of detection analysis proposed in~\cite{pagliaroli2}, an upper bound of 5.8~eV at 95\% C.L~\cite{pagliaroli1} was obtained. With a different likelihood approach and astrophysical parametrization of the neutrino emission, Lamb \& Loredo~\cite{ll} reached a 5.7~eV at 95\% C.L for the neutrino mass limit. 


There are good perspectives to explore the sub-eV region in the future experimental scenario for detection of galactic supernovae~\cite{pagliaroli1,nz1,nz2,nz3,janka}. Other mass limits obtained from the supernovae can be seen in the table presented in Ref.~\cite{pagliaroli3}.   

The perspective to detect $\nu_e$ from a supernova, but in a liquid scintillator detector, was recently considered in~\cite{scint1,scint2}. Water Cherenkov detectors with Gadolinium could also improve the number of $\nu_e$ events according to~\cite{gadolinium}. In the present work, we aim to discuss bounds on neutrino masses from supernovae motivated by the Zatsepin's idea. We are going to consider electronic neutrinos ($\nu_e$) emitted from the neutronization burst of a galactic supernova at a distance of 10~kpc and 15~M$_\odot$ and detected by a Liquid Argon Time Projection Chamber (LArTPC). 

The paper is organized as follow: Sec.~\ref{sn} presents the main assumptions about electronic neutrino production and its propagation; Sec.~\ref{detec} presents the basic features about the neutrino detection in the LArTPC; Sec.~\ref{method} presents the main assumptions about our event generated Monte Carlo simulation and Sec.~\ref{results} presents the boundaries on the neutrino mass. Sec.~\ref{discuss} and Sec.~\ref{conclu} present the discussion about our results and our main conclusions, respectively.  

\section{Neutrinos from Supernovae}\label{sn}

As the shock evolves against the infalling dense matter of the outer core of the star, some of its energy is lost by the photodissociation of nuclei into protons and neutrons. At this moment, there is an abundant production of electronic neutrinos ($\nu_e$) via $p+e^-\to n+\nu_e$. These electronic neutrinos accumulate in a very dense and opaque environment behind the shock wave. When the shock reaches the zone with densities around $10^{11}$g cm$^{-3}$ in a few miliseconds after the bounce the $\nu_e$'s are released. This process is called neutronization burst. It has a peak luminosity of $\sim 3.5\times10^{53}$~erg s$^{-1}$ and lasts $\sim 25$~miliseconds. For a recent review of the current knowledge and status of the explosion mechanism, see~\cite{janka_review}.    

At a distance $d$ of the progenitor star, for each flavor $\nu_{\beta}$ ($\beta=e,\bar e, x$), the unoscillated sprectral number fluxes ($F_{\nu_\beta}^0(E)$) are:
\begin{equation}
F_{\nu_\beta}^0(E)=\frac{L_{\nu_\beta}}{4\pi d^2}\frac{f_{\nu_\beta}(E)}{\langle E_{\nu_\beta} \rangle},
\label{eq4}
\end{equation}
where $L_{\nu_\beta}$ is the luminosity for the respective flavor $\nu_\beta$, $\langle E_{\nu_\beta} \rangle$ is the $\nu_\beta$ mean energy and $f_{\nu_\beta}(E)$ is the quasi-thermal spectrum written as follows~\cite{raffelt_spec}:
\begin{equation}
f_{\nu_\beta}(E)=\chi_\beta\left(\frac{E}{\langle E_{\nu_\beta} \rangle} \right)^{\alpha_\beta} e^{-(\alpha_\beta+1)E/\langle E_{\nu_\beta} \rangle}.
\label{eq5}
\end{equation}
The $\alpha_\beta$ is a parameter defined by $\langle E_{\nu_\beta}^2 \rangle/\langle E_{\nu_\beta} \rangle^2=(2+\alpha_\beta)/(1+\alpha_\beta)$ and $\chi_\beta$ is the normalization constant factor: $\int dE f_{\nu_\beta}(E)=1$. Several of these parameters, such as $L_{\nu_\beta}$ in Eq.~(\ref{eq4}), and $\langle E_{\nu_\beta}^2 \rangle$ and $\langle E_{\nu_\beta} \rangle^2$ in Eq.~(\ref{eq5}), change their respective values with time after bounce and depend on the simulation of the star explosion~\cite{garching}.


After taking into account the effect of neutrino oscillations, in the three neutrino families framework in a supernova environment, the fluxes ($F_{\nu_\beta}$) at Earth for each flavor $\nu_\beta$ are~\cite{amol}: 
\begin{equation}
F_{\nu_e}=pF^0_{\nu_e}+(1-p)F^0_{\nu_x},
\label{eq6}
\end{equation}
\begin{equation}
F_{\bar\nu_e}=\bar pF^0_{\bar\nu_e}+(1-\bar p)F^0_{\nu_x}
\label{eq7}
\end{equation}
and
\begin{equation}
4F_{\nu_x}=(1-p)F^0_{\nu_e}+(1-\bar p)F^0_{\bar\nu_e}+(2+p+\bar p)F^0_{\nu_x},
\label{eq8}
\end{equation} 
where $F^0_{\nu_\beta}$ are the primary unoscillated neutrino fluxes at the production region, $x=\mu$ or $\tau$ and $p$($\bar p$) is the survival probability of an electron (anti)neutrino after propagation through the SN mantle and the interestellar medium. For the actual oscillated parameters, $p=0$ for normal hierarchy (NH) ($m_1<m_2\ll m_3$) and $p=\sin^2\theta_{12}$ for inverted hierarchy (IH) ($m_3\ll m_2<m_1$). 

For massive neutrinos, the arrival time ($t_\oplus$) of the neutrino in the detector at Earth will be delayed~\cite{zatsepin}: 
\begin{equation}
t_\oplus=t_i + d/c +\Delta t,
\label{delay1}
\end{equation}
where $t_i$ is the emission time at the source and $c$ is the speed of light. The delay introduced by the neutrino mass $m_\nu$ in Eq.~(\ref{delay1}) is represented by $\Delta t$ and can be written as: 
\begin{equation}
\Delta t=\frac{d}{2c}\left(\frac{m_\nu}{E}\right)^2.
\label{delay2}
\end{equation}

\section{Detection Assumptions}\label{detec}

The neutrino flux of the neutronization burst is mainly populated by electronic neutrinos. Therefore, only experimental technniques with high sensitivity to $\nu_e$ can have a possibility to investigate this particular phase of supernova explosion. 

The Liquid Argon Time Projection Chamber (LArTPC), such as the ICARUS-like detectors~\cite{icarus}, have all the features required to identify the neutronization burst. The ``Long Baseline Neutrino Experiment'' (LBNE)~\cite{lbne} foresees a 34~kton (fidutial volume) LArTPC as its far detector. The experiment is dedicated for determination of CP violation phase and neutrino mass hierarchy. Though, its dimensions and underground site are suitable for exploring rare events, such as neutrinos from supernovae and proton decay. This experimental design is taken into account for our calculations.




The most favourable electronic neutrinos ($\nu_e$) detection channel in LArTPCs is the charged current (CC) interaction: $\nu_e+^{40}Ar\to e^- +^{40}K^*$. There are other detection channels, but all of them with smaller cross section. These detection channels can be tagged by spectral analysis of the photon emission from de-excitation of $K$, $Cl$ or $Ar$, which exhibit specific spectral lines. The $\nu-Ar$ cross sections can be found in Fig.~3 of Ref.~\cite{rubbia_ar}. 

\section{Calculation Method}\label{method}

The number of expected events in the detector can be evaluated in the following way:
\begin{equation}
\frac{dN}{dt}(t_\oplus)=\frac{n_t}{4\pi d^2}\int dE \sigma(E)\int dt_i F_{\nu_\beta}(E,t_i)\delta (t_\oplus-t_i-\Delta t),
\label{eq3}
\end{equation}
where $\sigma(E)$ is the $\nu_e-Ar$ CC cross section, $n_t$ is the number of targets in the 34~kton Argon detector and $F_{\nu_\beta}(E,t_i)$ is the oscillated neutrino flux, which is evaluated by Eqs.~(\ref{eq4}-\ref{eq8}). 

We performed a Monte Carlo simulation (MC) with very optimistic assumptions. Nevertheless, all of our assumptions have solid grounds and can be taken as a ``realistic case''. The number of expected events depends on the considerations we make about the neutrino production and their propagation till the detector. We considered a comprehensive knowledge of the astrophysical process related to the SN explosion, which means that the astrophysical parameters related to neutrino production, ie the neutrino fluxes, are well known. However, to introduce uncertainties related to the explosion mechanism, we considered a 5\% error in the overall normalization of the evaluated counting rates. Concerning the SN distance, we assumed a galactic SN of 10~kpc. The detector configuration, as pointed out before, is 34~kton LArTPC with 80\% of efficiency for $\nu_e$ charged current (CC) detection. For now, background effects have been disconsidered since the detector is assumed to be placed underground. In this situation we expect a suppression of seven orders of magnitude~\cite{internal}. The $\nu_e$ energy threshold for detection was set to 5~MeV. For the energy resolution we used a parametrization given by: $11\%/\sqrt{E(MeV)}+2\%$~\cite{ener_resolution}. For the time resolution, we adopted a $\pm$20~$\mu s$ flat distribution. Also, we stress the fact that we performed our MC ignoring aditional complications given by the lack of knowledge if the first detected event is syncronized or not with the beginning of the burst. In the MC we included the two possible scenarions for mass hierarchy, as mentioned in Sec.~\ref{sn}, with the best-fit values of the oscillation parameters with no CP violation effects~\cite{osc}.

\section{Results}\label{results}

\begin{figure}[!h]
     \centering
     \includegraphics[scale=0.35]{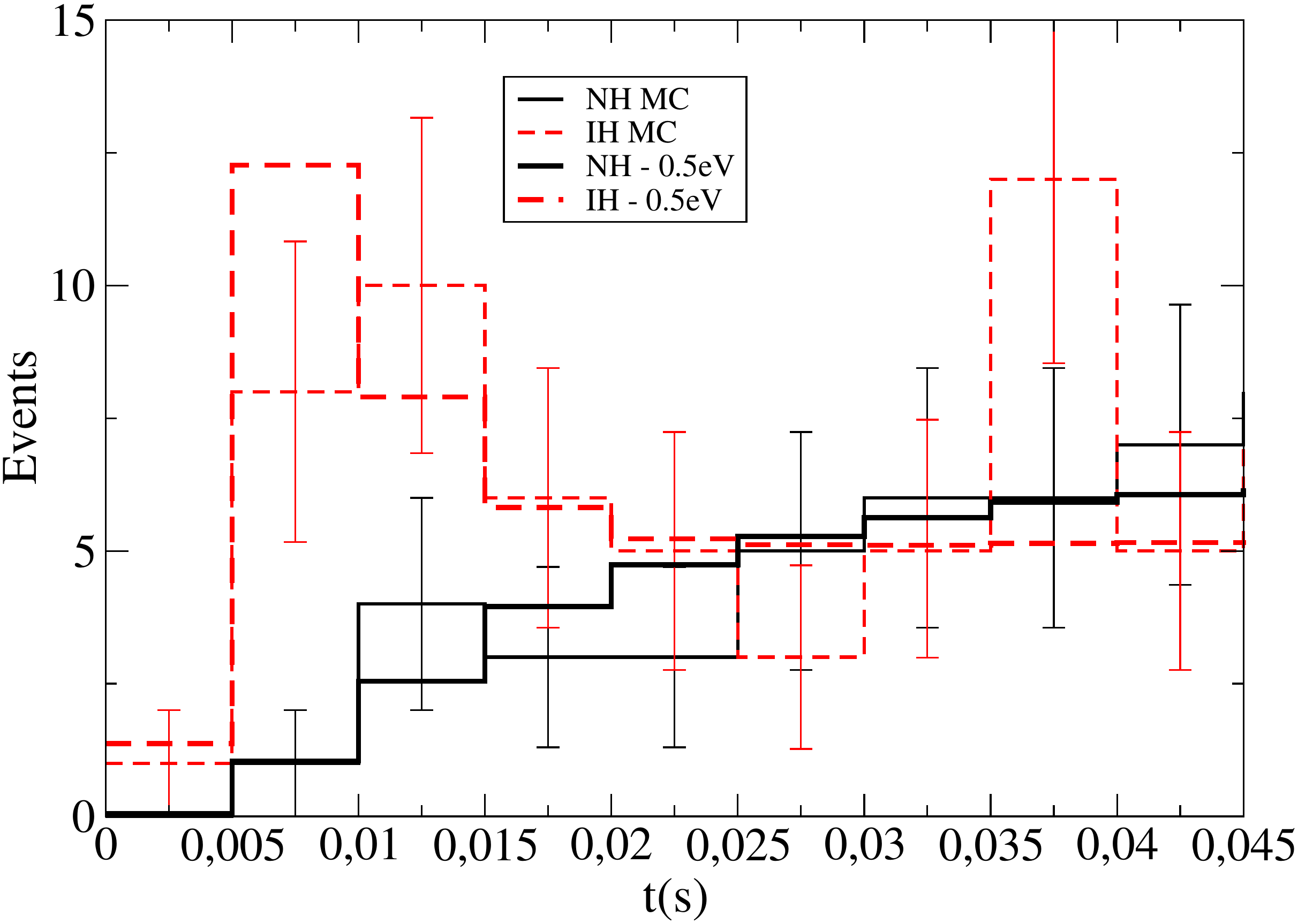}
     \caption{{\it (Colours online)} Results of our Monte Carlo realization. For further details see text. Normal hierarchy (NH) in solid curves and inverted hierarchy (IH) in dashed curves. Thicker lines represent the expected number of events for a 0.5~eV neutrino mass.}
     \label{MC}
\end{figure}

Our MC simulation results in thinner lines are shown in Fig.~\ref{MC}. NH and IH hierarchy are represented by solid and dashed lines, respectively. The error bars in each bin are assumed to follow a Poissonian distribution and each bin has $\delta t=0.005$~s width. In Fig.~\ref{MC}, we also show the expected number of events in the thicker lines considering a 0.5~eV neutrino mass, just for comparison with our MC.

Non vanishing neutrino mass causes a delay in the experimental signal, changing the expected number of events in each time bin, as we can notice analysing Eq.~(\ref{delay2}) and Eq.~(\ref{eq3}). These theoretical modifications can be compared with the number of events in each bin generated by our MC simulation using the $\chi^2$ test:
\begin{equation}
\chi^2=\sum_{i=1}^N\frac{(N^{MC}_i-N_i(m_\nu))^2}{\sigma_i^2},
\label{eq:chi}
\end{equation}
where $i=1,N$ bins, $N_i^{MC}$ is the number of events generated by our MC in the $i$-bin, $N_i(m_\nu)$ is the number of events evaluated by Eq.~(\ref{delay2}) and Eq.~(\ref{eq3}) for different neutrino masses ($m_\nu$) and $\sigma_i$ is the associated error. 

\begin{figure}[!h]
     \centering
     \includegraphics[scale=0.7]{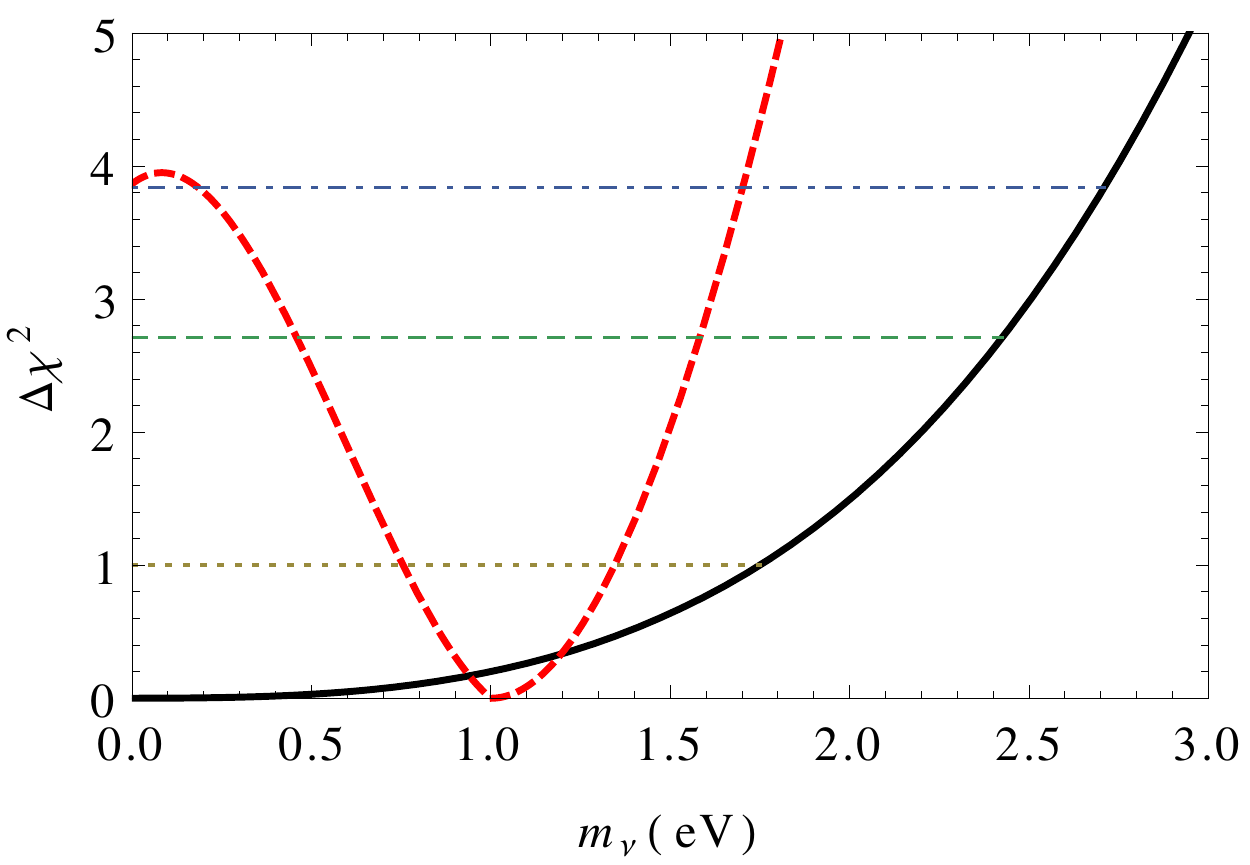}
     \caption{{\it (Colours online)} Neutrino mass bound considering a supernova at 10~kpc and a liquid argon detector with 34~kton, such as the one projected in the LBNE experiment. Dotted line is 1$\sigma$ limit; dashed line is the 2$\sigma$ limit; and dotted-dashed line is the 3$\sigma$ limit. The strong red dashed curve is for IH and the strong solid black curve is for NH.}
     \label{fig1}
\end{figure}

The ``realistic case'' provided $\chi^2_{min}\approx2.12$ for a neutrino mass of $m_\nu=0.00$~eV and considering NH. From the $\Delta\chi^2=\chi^2(m_\nu)-\chi^2_{min}$, the upper bound on neutrino mass from the $\nu_e$ neutronization burst stage is, at 95\% C.L, 2.71~eV. On the other hand, for the IH case, we have obtained $\chi^2_{min}\sim 5.43$ for a neutrino mass of $m_\nu\approx1.00$~eV. At 95\% C.L, $0.18~\mbox{eV} \lesssim m_\nu \lesssim 1.70$~eV. We present our $\Delta \chi^2$ in terms of the neutrinos mass, $m_\nu$(eV), in Fig.~\ref{fig1}, for both hierarchies. In these plots, dotted lines are bounds with 68.27\% C.L, dashed lines are for 90\% C.L; and dotted-dashed lines are the 95\% C.L. 

We also performed other MC simulations and did the $\chi^2$ analysis for the mass as a parameter for other situations that are described in the following: 

\begin{enumerate}

\item ``perfect case'': we consider that the normalization is completely known, 100\% of detector efficiency and a perfect time and energy resolution. We mantain an energy cutoff in 5~MeV;


\item ``perfect norm'': returning to the conditions estabilished in ``realistic case'', but now for a completely known supernova normalization; 


\item ``perfect eff'': considering a detector with 100\% of efficiency and other conditions of the ``realistic case'' maintained; 


\item ``cut ener'': keeping the ``realistic case'', but instead cutting events with energy less than 5~MeV, we perform a energy cut at $\approx 2.2$~MeV; 


\item ``time resol'': considering a temporal resolution of 1~ms instead of 20~$\mu s$ of the ``realistic case''; 


\item ``bin1'': if we increase the bin size from 5~ms to 10~ms using the ``realistic case'' and


\item ``bin2'': if we decrease the bin size from 5~ms to 2~ms using the other conditions of the ``realistic case'' 


\end{enumerate}

We present in Table~\ref{tabresumonh} our mass boundaries at 68.27$\%$, 90$\%$ and 95$\%$ C.L for NH. The numbers in parentheses are the boundaries for IH.  In this table we also show the minimum value of the mass ($m_{min}$) with its respective $\chi^2$, the $\chi^2_{min}$. 

\begin{table}[h!]
\centering
    \begin{tabular}{| l | l | l | l | l | l |}
    \hline
      & 68.27$\%$ C.L &  90$\%$ C.L &  95$\%$ C.L & $m_{min}$ & $\chi^2_{min}$ \\ 
    \hline
    realistic & 1.75(0.76-1.35) & 2.43(0.46-1.58) & 2.71(0.18-1.70) & 0.00 (1.00) & 2.12 (5.43)  \\ 
   \hline
    perfect norm & 0.91-3.25(0.65) & 3.79(0.95) & 4.07(1.1) & 2.27 (0.00) & 8.51 (3.22) \\
   \hline
    perfect eff & 2.01(0.82) & 2.52(1.08) & 2.76(1.20) & 1.10 (0.00) & 2.39 (3.06) \\
   \hline
    cut ener & 2.05(0.69) & 2.58(1.03) & 2.87(1.20) & 1.22 (0.00) & 7.53 (6.34) \\
   \hline
    time resol & 1.34(0.71) & 1.94(1.03) & 2.22(1.17) & 0.00 (0.00) & 2.33 (8.53)\\
   \hline
    bin1 & 0.71-2.23(0.63) & 2.66(0.93) & 2.88(1.09) & 1.54 (0.00) & 0.51 (3.22)\\
   \hline  
    bin2 & 1.65(2.24-2.67) & 2.14(2.03-2.91) & 2.38(1.90-3.06) & 0.81 (2.50) & 4.84 (9.74)\\
   \hline
    perfect & 3.26(0.96) & 3.74(1.19) & 3.96(1.29) & 2.37 (0.50) & 12.20 (4.41)\\	
   \hline
    \end{tabular}
\caption{Upper Bounds on $m_\nu$(eV) considering a supernova at 10~kpc and a liquid argon detector with 34~kton. We show the results for normal hierarchy (NH) and the numbers inside the parentheses are the mass evaluation for inverted hierarchy (IH). Also we show the the minimum value of the mass ($m_{min}$) with its respective $\chi^2$, the $\chi^2_{min}$. For details of the meaning of each test, see the text.}
\label{tabresumonh}
\end{table}

\section{Discussion}\label{discuss}

From Table~\ref{tabresumonh} we notice that, despite we change the configuration of the detector, statistical fluctuation of the Monte Carlo simulation plays a fundamental role in the mass bound determination, since we have few events in the neutronization burst and this process lasts only $\approx$~25~ms. A very good example of this kind of behaviour is the one shown in the IH curve of Fig.~\ref{MC} (red dashed). In the interval of 35-40~ms, there is a peak that was not theoreticaly predicted. So, this generates a worse $\chi^2$ and an interval in the neutrino mass bound (0.18-1.70~eV) with a $\chi^2_{min}$ at 1.00~eV. Another example of the impact of this statistical fluctuation is that, for the NH case, we have obtained a more stringent bound in the ``realistic'' case compared to the ``perfect'' case. 

Our bound, at 95\% C.L, for the NH is very similar to the one obtained by the Troitsk experiment of tritium $\beta$ decay. So, if we really have NH and comparing with terrestrial experiments, which are independent of mass hierarchy, like Katrin ($m_\nu<0.2$~eV), it will be very difficult that LBNE will put a more stringent limit on neutrino mass. From the Planck Cosmic Microwave Background measurements and galaxy clustering information from the Baryon Oscillation Spectroscopic Survey (BOSS), part of the Sloan Digital Sky Survey-III, and assuming a $\Lambda$CDM cosmological model, Ref.~\cite{cosmo1} obtained, at 95\% C.L, $\sum m_\nu<0.35$~eV. Ref.~\cite{cosmo2} obtained that $\sum m_\nu<0.18$~eV, when one takes into account observations of the large-scale matter power spectrum from the WiggleZ Dark Energy Survey, Planck sattelite and baryon acoustic oscillation. If we consider that mass eigenstates are degenerated, $m_\nu\lesssim 0.12$~eV~\cite{cosmo1} and $m_\nu\lesssim 0.06$~eV~\cite{cosmo2}. 

For the NH and IH case, our bounds, respectively, at 95\% C.L, are $m_\nu<2.71$~eV and $0.18~\mbox{eV}<m_\nu<1.70$~eV. These bounds are almost similar with other prospects of neutrino mass bounds obtained for SN neutrinos. Just for comparison, authors in~\cite{mass_juno} considering only normal hierarchy, using a 20~kton liquid scintillator at Jiangmen Underground Neutrino Observatory (JUNO) and a galatic supernova $\bar \nu_e$ signal obtained a limit on the absolute neutrino mass scale: $m_\nu < (0.83\pm 0.24)$~eV at 95\% C.L. This limit is also similar with the ones obtained in the other cases discussed in this work - see Table~\ref{tabresumonh}. So, our conclusion here is that, for a SN at 10~kpc of distance and 15~M$_\odot$, LBNE 34~kton argon detector can probe neutrino masses of order of 1~eV, at 95\%~C.L. We stress the fact that knowing neutrino mass hierarchy is fundamental to put limits on neutrino masses from SN and LBNE has as one of its objectives determine this property.

\section{Conclusions}\label{conclu}      

Even though our supernova neutrino bound seems not to be competitive with the prospective Katrin bound, the neutrino mass bound from SN has a lot merit and importance, and, of course, it should be a source of strong investigation. With improvements in the detector efficiency, the number of events per bin would increase and a more competitive bound could be obtained. We remember that efficiency includes several effects, such as, loss of events due to light attenuation, fluctuations of the number of photoelectrons, the geometry of the detector etc. These are still unknown sources of errors, and their knowledge is crutial for a more detailed work than presented here. We are aware that our assumptions are very crude, however, as stated by the LBNE collaboration in~\cite{lbne}, work in understanding the physics and choices of detector is still underway. It is necessary to determine, for example, the low energy response of the detector, its geometry, its real threshold, the reconstruction of the events, event vertex resolution, the energy resolution, absolute event $ms$ time precision and other details. That is why we are being very conservative, but at the same time, very optimistic in this work. However, we should stress that considering this time binned analysis, we have demonstrated that statistical fluctuation may play a major role in the determination of the neutrino mass bound. We point out that a more dedicated MC simulation, with real and proper detector information, with an effective study of a possible event-by-event likelihood of detection for this flux of $\nu_e$ of neutronization burst and, finally, the importance of the time of the first event detected should be studied deeply in the foreseeable future. 

\section*{Acknowledgments}
F. Rossi-Torres would like to thank CNPq for the full financial support. E.~Kemp would like to thank FAPESP and CNPq for partial financial suppport. M. M. Guzzo also would like to thank CNPq for partial financial support to this work. Also, we would like to thank prof.~H.-T.~Janka for providing the data of supernova simulation used to calculate neutrino fluxes. 










\end{document}